\documentclass[journal]{IEEEtran}

\usepackage{url}
\usepackage{tabularx}
\usepackage{graphicx}
\usepackage[cmex10]{amsmath}
\usepackage{subfigure}
\usepackage{cite}
\usepackage{float}
\usepackage{color}
\usepackage{soul} 
\usepackage{flushend}
\usepackage{amsfonts}
\usepackage{amssymb}

\ifCLASSINFOpdf
\else
\fi
\graphicspath{{figures/}}

\def\min{\mbox{min}}
\def\max{\mbox{max}}
\def\rea{\mathbb{R}}

\def\begequarrs{\begin{eqnarray*}}
	\def\endequarrs{\end{eqnarray*}}
\def\begequarr{\begin{eqnarray}}
\def\endequarr{\end{eqnarray}}
\def\begarr{\begin{array}}
	\def\endarr{\end{array}}
\def\begequ{\begin{equation}}
\def\endequ{\end{equation}}
\def\lab{\label}
\def\begdes{\begin{description}}
	\def\enddes{\end{description}}
\def\begenu{\begin{enumerate}}
	\def\begite{\begin{itemize}}
		\def\endite{\end{itemize}}
	\def\endenu{\end{enumerate}}

\def\lef[{\left[\begin{array}}
	\def\rig]{\end{array}\right]}
\def\qed{\hfill$\Box \Box \Box$}
\def\begcen{\begin{center}}
	\def\endcen{\end{center}}
\def\begrem{\begin{remark}\rm}
	\def\endrem{\end{remark}}
\def\begcas{\begin{cases}}
	\def\endcas{\end{cases}}

\usepackage{color}
\newcommand{\ro}[1]{{\color{black} #1}}
\definecolor{myblue}{RGB}{0,0,153}

\newtheorem{assumption}{Assumption}
\newtheorem{proposition}{Proposition}

\newtheorem{remark}{Remark}
\begin{document}

\title{Voltage regulation in buck--boost coniverters\\ feeding an unknown constant power load:\\ an adaptive passivity--based control}

\author{Carlos A. Soriano--Rangel, Wei He, Fernando Mancilla--David, \IEEEmembership{Member, IEEE}, and Romeo Ortega, \IEEEmembership{Fellow, IEEE}
\thanks{Manuscript submitted November 29, 2018. Corresponding author: Wei He.}
\thanks{C. A. Soriano--Rangel and F. Mancilla--David are with the University of Colorado Denver, Campus Box 110, P.O. Box 173364 Denver, CO 80217--3364, USA, (e--mail: [carlos.sorianorangel]fernando.mancilla-david@ucdenver.edu).}
\thanks{W. He (corresponding author) is with  School of Automation, Nanjing University of
Information Science and Technology, Ningliu road No. 219, 210044, Nanjing, China, (e--mail: hwei@nuist.edu.cn).}
\thanks{R. Ortega is with LSS/CNRS/CentraleSup\'elec, 3, Rue Joliot--Curie 91192 Gif--Sur--Yvette, France, (e--mail: romeo.ortega@lss.supelec.fr). R. Ortega is also with the ITMO University, Kronverkskiy Prospekt, 49, Sankt-Peterburg, Russia, 197101.}
}


\maketitle

\begin{abstract}
Rapid developments in power distribution systems and renewable energy have widened the applications of dc--dc buck--boost converters in dc voltage regulation. Applications include vehicular power systems, renewable energy sources that generate power at a low voltage, and dc microgrids. It is noted that the cascade--connection of converters in these applications may cause instability due to the fact that converters acting as loads have a constant power load (CPL) behavior. In this paper, the output voltage regulation problem of a buck--boost converter feeding a CPL is addressed. The construction of the feedback controller is based on the interconnection and damping assignment control technique. Additionally, an immersion and invariance parameter estimator is proposed to compute online the extracted load power, which is difficult to measure in practical applications. It is ensured through the design that the desired operating point is (locally) asymptotically stable with a guaranteed domain of attraction. The approach is validated via computer simulations and experimental prototyping.
\end{abstract}



\section{Introduction}
{\color{black} Nowadays, the DC--DC buck--boost converter is being broadly adopted in vehicular power systems (e.g., sea, land, air and space vehicles); unconventional energy systems (e.g., photovoltaic panels, fuel cells, piezoelectric); and dc microgrids, due to its voltage step--up and step--down capabilities \cite{khalig08,lefeuvre07,pavlovic14,martin15}.} For instance, the architecture of grid--connected double--stage photovoltaic power system may include the cascade connection of a photovoltaic array, a dc--dc converter and an inverter \cite{martin15}.

The converter acting as a load (the inverter in the previous example) is often controlled to synthesize a certain amount of power and therefore will exhibit a constant power load (CPL) behavior. As compared to a passive load where the voltage--current relationship is restricted to the first and third quadrants, CPLs correspond to hyperbolas in this space \cite{kwa11,kar17}. Because of this, the existence of CPLs may affect the dynamic behavior of the power system and even could induce erratic or unstable behavior \cite{Solsona15,Du13}.

Although the control of these converters in the face of classical loads is well understood \cite{geyer08,son12,kim14,wang16,wang17}, the proliferation of CPLs  poses a new challenge to control theorists \cite{Bar16,Mar12,Sin17,emadi06}. It is noteworthy that the potentially large variations of the operating point caused by the varying input voltage render linear approximations inadequate, and in order to capture the complete dynamics a nonlinear model is required. Passivity--based controllers (PBCs) have been applied to stabilize this type of systems with switching devices \cite{konst13,meshram18}. Most of the aforementioned applications require to regulate the output voltage at a predetermined level while being fed by energy sources that generate power within a wide voltage range need to step--up and step--down the input voltage depending on the operating point. Buck--boost converters provide such capability.

Several approaches can be found in recent literature for voltage regulation of the buck--boost converter with a CPL, such as passive damping \cite{Cespedes11}, feedback linearization \cite{Rah10,Salimi16}, active--damping \cite{Rah09}, sliding--mode control \cite{Sin16}, and the pulse adjustment method \cite{khalig08}. It should be noted that although the aforementioned techniques have addressed this problem, they do not provide the guaranteed stability properties for the original nonlinear system.

\ro{The recent research presented in \cite{He2018} proposes an adaptive PBC and provides a complete stability analysis for a buck--boost converter feeding a CPL. The approach combines interconnection and damping assignment (IDA) control \cite{Ortega02} and the immersion and invariance (I\&I) technique for estimation of unknown parameter \cite{Ast08}. However, the control law of \cite{He2018} is provided in terms of a time--scaled model and is extremely complicated to be of practical interest. A second approach, presented in \cite{He2018CEP,ACC2018}, addresses the same control problem, but synthesizes a significantly simpler control law. The key modification that leads to simplifying the control law is a partial linearization that transforms the model into a cascade form. Nevertheless, the approach still relies on the same time--scaled model, which is hard to deal with in practical applications.

The approach presented herein overcomes the aforementioned limitations, proposing an adaptive PBC simple enough to be implemented in practice, with no time scaling or any kind of linearization techniques. The specific contributions of this paper are:
\begin{itemize}
  \item Proposition of an adaptive PBC to stabilize a DC--DC buck--boost converter feeding a CPL without using any time scaling or linearization methods
  \item A complete stability analysis of the closed--loop system under the proposed controller
  \item Experimental validation, including reference tracking, as well as line and load regulation
\end{itemize}}

The remainder of the paper is organized as follows. Section \ref{section2} contains the model of the system and problem formulation. Section \ref{section3} presents the adaptive IDA--PBC. Simulation and experimental results are provided in Section \ref{section4}. Finally, the concluding remarks of Section \ref{section5} close the paper.

\section{System model and control problem formulation}
\label{section2}
 \subsection{Model of buck--boost converter with a CPL}
\label{subsec21}

\begin{figure}[H]
  \centering\includegraphics[scale=1]{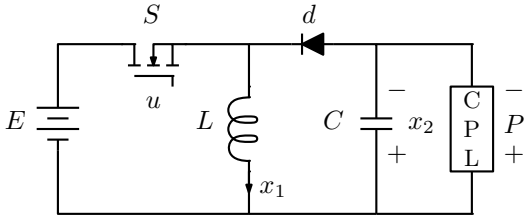}
  \caption{Circuit schematic of the buck--boost converter feeding a CPL.}\label{buckboostcircuit}
\end{figure}

The circuit schematic of a buck--boost converter feeding a CPL is shown in Fig. \ref{buckboostcircuit}. Under the standard assumption that it operates in continuous conduction mode, the average model is given by
\begin{align}
\nonumber
L {\frac{dx_1}{dt}}&= -(1-u)x_2 + uE, \\
C {\frac{dx_2}{dt}}&= (1-u)x_1-\frac{P}{x_2},
\label{bucboo}
\end{align}
where $x_1 \in \mathbb{R}{>0}$ is the inductor current, $x_2 \in \mathbb{R}{>0}$ the output voltage, $P\in \mathbb{R}{>0}$ the power extracted by the CPL, $E\in \mathbb{R}{>0}$ is the input voltage and $u\in [0,1]$ is the duty ratio of the switch $S$, which is the control signal.

The assignable equilibrium set of the system is given by
\begin{align}
\label{set1}
\mathcal{E}:=\left\{ (x_1,~x_2) \in \mathbb{R}^2_{>0}\;|\; x_1- P\left(\frac{1}{x_2}+ \frac{1}{E}\right)=0  \right\}.
\end{align}

\subsection{Control problem formulation}
\label{subsec22}

The control problem is formulated assuming the following about the system described by \eqref{bucboo}:

\begin{assumption}
\lab{ass1}
 The power load ($P$) is unknown, while the parameters $L$, $C$ and $E$ are known.
\end{assumption}

\begin{assumption}
\lab{ass2}
The state $(x_1,~x_2)$ is measurable.
\end{assumption}

The control problem is to design a state--feedback control law where:

\begin{itemize}
\item  $x_\star=(x_{1\star},x_{2\star})$  is an asymptotically stable equilibrium of the closed--loop with a well--defined domain of attraction
\item It is possible to define an invariant set of initial conditions $\Omega \subset \rea^2_{>0}$, where $x(0)$ in $\Omega$ implies $x(t)$ in $\rea^2_{>0}$ and $x(t)$ to $x_\star$.
\end{itemize}

Given that the state to control is $x_2$, a reference $x_{2\star}$ is fixed and then $x_{1\star}$ is calculated using \eqref{set1}.

\section{Proposed control scheme}
\label{section3}
Following a similar approach as that of \cite{He2018CEP}, the controller design proceeds in two steps:

\begenu
\item Applying the IDA--PBC method to stabilize the system by assuming $P$ known and ensure local stability of the desired operating point
\item Designing an I\&I estimator for the power load such that the above scheme is adaptive
\endenu
\subsection{IDA--PBC}

In this subsection, the control law obtained through the IDA--PBC method is presented. In order to avoid notation cluttering, the voltage across the switch throw ($v_\mathrm{T}=E+x_2$) and a linear combination of the capacitor and inductor energies ($W=Cx_2^2+2Lx_1^2$) have been introduced, as well as $W_\star=Cx_{2\star}^2+2Lx_{1\star}^2$.

\begin{proposition}%
\label{proposition0}%
The IDA--PBC law given by

\begin{align}
\label{ubucboo}
u =& \frac{1}{\frac{x_1^2}{C^2}+\frac{v_\mathrm{T}^2}{L^2}}\Bigg(\frac{x_1}{C^2}\left(x_1 - \frac{P}{x_2}\right)+\frac{x_2 v_\mathrm{T}}{L^2}-\left(\frac{2 x_1 x_2}{C^2 v_\mathrm{T}}+ \right. \nonumber \\
& \left.\frac{x_2 v_\mathrm{T}}{L^2 x_1}\right)\bigg(\frac{1}{C^2 W^2} \bigg(\sqrt{2LC^{6}W}EP  x_1 \arctan\left(\frac{\sqrt{2 L} x_1}{\sqrt{W}}\right) \nonumber \\
& -C^3 P W v_\mathrm{T}+k_1 L x_1 W^2 \left(W + 2 C k_2\right)\bigg)\bigg) \nonumber \\
& + \frac{1}{2 L C x_2 W^2}\left(\frac{2 L E x_1^2}{C^3 v_\mathrm{T}^2}-\frac{2 x_2}{LC}\right) \bigg(\sqrt{2LC^6W} EP x_2^2 \nonumber \\
& \arctan \left(\frac{\sqrt{2 L} x_1}{\sqrt{W}}\right)+ C^2 W \left(2 L P x_1 v_\mathrm{T}-E x_2W\right)+ \nonumber \\
& k_1 L \left(x_2W\right)^2 \left(W + 2 C k_2\right)\bigg)\Bigg),
\end{align}

ensures that:

\begin{itemize}
\item $x_\star$ is an asymptotically stable equilibrium of the closed--loop system with Lyapunov function

\begin{align}
\label{solution} \hspace{-1.3cm}
H_d(x)&=-\frac{\sqrt{C}}{2L}\left(\sqrt{C}Ex_2+\sqrt{2L}P\arctan{\left[\frac{\sqrt{2L}x_1}{\sqrt{C}x_2}\right]}\right) \nonumber \\
& -\frac{\sqrt{2L}PE\arctan{\left[\frac{\sqrt{2L}x_1}{\sqrt{W}}\right]}}{\sqrt{\frac{W}{C}}} \nonumber \\
& +\frac{k_1}{4C}\left(W + 2 C k_2\right)^2.
\end{align}

\item  There exists a positive constant $c$ such that the sublevel sets of the function $H_d(x)$
 \begin{align}\label{region}
  \Omega_x:=\{x \in \mathbb{R}_{>0}^{2}\;|\; H_d(x)\leq c\},
  \end{align}
are an estimate of the domain of attraction ensuring the state trajectories {\em remain} in $\mathbb{R}_{>0}^{2}$. That is, for all $x(0)\subset \Omega_x$, $x(t) \subset \Omega_x,\forall t \geq 0,$ and $\lim_{t \to \infty}x(t)=x_\star$.
\end{itemize}
\end{proposition}
\noindent $k_1$ is a tuning gain that needs to satisfy the following condition:

\begequ
\lab{bouk1}
k_1  > \max \{k_1', k_1''\}.
\endequ

$k_1', k_1''$ are defined with more detail in the full version of the paper. Constant $k_2$ is then calculated as

\begin{align}
\lab{k2}
k_2:=&\frac{1}{2 C k_1 L x_{1\star} W_\star^2}\Bigg(\sqrt{2LC^6W_\star} EP x_{1\star} \arctan\left(\frac{\sqrt{2L}x_{1\star}}{\sqrt{W_\star}}\right) \nonumber\\
& -W_\star \bigg(C^3 P (E -x_{2\star})+C^2 k_1 L x_{1\star} x_{2\star}^4+ \nonumber \\
& 4 C k_1 L^2 x_{1\star}^3 x_{2\star}^2+4 k_1 L^3 x_{1\star}^5\bigg)\Bigg).
\end{align}

\textit{Proof:} The proof is shown in Section \ref{app1} of Appendix.  

\subsection{I\&I power estimator}
In this subsection, the fact that $P$ is usually unknown is addressed via an I$\&$I estimator.

\begin{proposition}\label{proposition2}
	Consider the buck--boost converter of \eqref{bucboo} satisfying Assumptions \ref{ass1} and \ref{ass2} in closed--loop with an adaptive version of the control \eqref{ubucboo} given by
\begequ
\lab{adacon}
u = \bar u(x,\hat P,k_1)
\endequ
where $\hat P(t)$ is an on--line estimate of $P$ generated with the I$\&$I estimator
	\begin{align}\lab{boest1}
	\hat P=& -\frac{1}{2}\gamma C x_2^2+{P}_I\\
	\dot{P}_I=&\gamma x_1x_2 (1-u)+ \frac{1}{2}\gamma^2Cx_2^2 -\gamma P_I \label{boest2}
	\end{align}
where $\gamma>0$ is a free gain. There exists $k_1^{\min}$ such that for all $k_1>k_1^{\min}$ the overall system has an asymptotically stable equilibrium at $(x, \hat P)=(x_\star, P)$.

The proof of power load estimator \eqref{boest1}, \eqref{boest2} is given in Section \ref{app2} of Appendix.
To prove asymptotic stability of $(x,\hat P)=(x_\star,P)$ the adaptive controller \eqref{adacon} is written as
$$
\bar u(x,\hat P,k_1) = \bar u(x, P, k_1) + \delta(x, \tilde P,k_1),
$$
where the mapping
$$
\delta(x,\tilde P,k_1) := \bar u(x,\tilde P + P, k_1)- \bar u(x,P,k_1),
$$
has been defined. It is noteworthy that $\delta(x,0,k_1)=0.$

Invoking the proof of proposition \ref{proposition0} the closed--loop system is now a cascaded system of the form
\begequarrs
\dot x & = & F_d(x) \nabla H_d(x)+g(x)\delta(x,\tilde P,k_1) \\
\dot {\tilde P}&= &-\gamma \tilde P,
\endequarrs
where $g(x)$ is the system input matrix
\begequ
\lab{g}
g(x):=\left[
  \begin{array}{c}
    C(x_2+E) \\
    -Lx_1. \\
  \end{array}
\right]
\endequ

Now, $\tilde P(t)$ tends to zero exponentially fast for all initial conditions, and for sufficiently large $k_1$, {\em i.e.}, such that \eqref{bouk1} is satisfied, the system above with $\tilde P=0$ is asymptotically stable. Invoking well--known results of asymptotic stability of cascaded systems, {\em e.g.}, Proposition 4.1 of \cite{Sep97}, completes the proof of (local) asymptotic stability.

\end{proposition}

\section{Computer simulations and experiments}
\lab{section4}
This section validates the theoretical results of Section \ref{section3} via computer simulations and experimental prototyping. Computer simulations are implemented in MATLAB/Simulink release R2017b. \ro{The prototyping of the buck--boost converter feeding a CPL is realized using commercial off--the--shelf Vishay Dale converter boards model MPCA75136 and a Texas Instrument DSP model TMS320F28335. Table \ref{tab:simpars} summarizes simulation and experimental set--points utilized as case studies, along with the physical parameters of the MPCA75136 boards. It is noted that simulations and experiments are performed using the same system characterization, and therefore results are directly comparable.}

\begin{table}[!ht]
\caption{Simulation/experimental setpoints and physical parameters.}
\centering
\begin{tabular}{lccc}
\hline
~Parameter~&Symbol (unit)~~& \multicolumn{2}{c}{Value} \\
~ & ~ & Boost & Buck\\
\hline
~Input voltage~~ & $E~(\text{V})$~~~ & $15$ & $15$  \\
~Reference output voltage~~ & $x_{2\star}(\text{V})$~~ & $25$ & $12$  \\
~Gain~~ & $x_{2\star}/E$~~ & $1.67$ & $0.8$  \\
~Nominal extracted power~~ & $P~(\text{W})$~~~ & ${20,~30}$ & ${6,~12}$\\
~Inductance~~ & $L,L'~(\mu \text{H})$~~~ & 216.8 & 216.8 \\
~Capacitance~~ & $C, C'~(\mu \text{F})$~~~ & 1380 & 1380\\
\hline
\end{tabular}
\label{tab:simpars}
\end{table}

Several tests of the closed loop system with the proposed controller are done using averaged and switched simulations, and experiments. The average system of the circuit shown in Fig. \ref{buckboostcircuit} in closed loop with the IDA--PBC of \eqref{ubucboo} and the power estimator of \eqref{boest1}--\eqref{boest2} are simulated and used to perform a gain sensitivity analysis, and to obtain the phase plots of the system. Results of simulations using different values for $k_1$ and $\gamma$ are used for the gain sensitivity analysis, while the phase plots are obtained by running the closed loop simulation with different initial values for $x_1$ and $x_2$. \ro{Besides, a performance comparison of the proposed adaptive PBC aginst a traditional PI controller is presented.}

Then, the controller's ability to regulate the output voltage when a step is applied to the extracted power is evaluated experimentally and compared with simulation results for a chosen set of gains. To this end, a switched simulation of the closed loop system is implemented in MATLAB/Simulink's Simscape electrical toolbox. In this simulation, additional components such as sensing resistors which are installed in the converters are taken into consideration to match closer the actual converter boards.

Finally, the ability of the IDA--PBC to regulate the output at the desired voltage when the input voltage changes (line regulation) and when the load changes (load regulation) is tested experimentally.

\subsection{Averaged simulations}
\subsubsection{Gain sensitivity analysis}
An averaged simulation of a buck--boost converter feeding a CPL \eqref{bucboo} with the proposed IDA--PBC \eqref{ubucboo} while operating in boost mode is carried out. The simulation parameters are prsented in Table \ref{tab:simpars}. Herein, a step change is applied to $P$ and the transient profiles of both states is observed for different values of the control gain $k_1$, while keeping $\gamma$ constant at 20. Then, the estimator's transient performance is evaluated for different values of $\gamma$ while keeping $k_1$ constant at 0.1. The results of this analysis are taken then into consideration to select proper gains for the experiments.

\begin{figure}[!ht]
\centering
\includegraphics[width=\columnwidth]{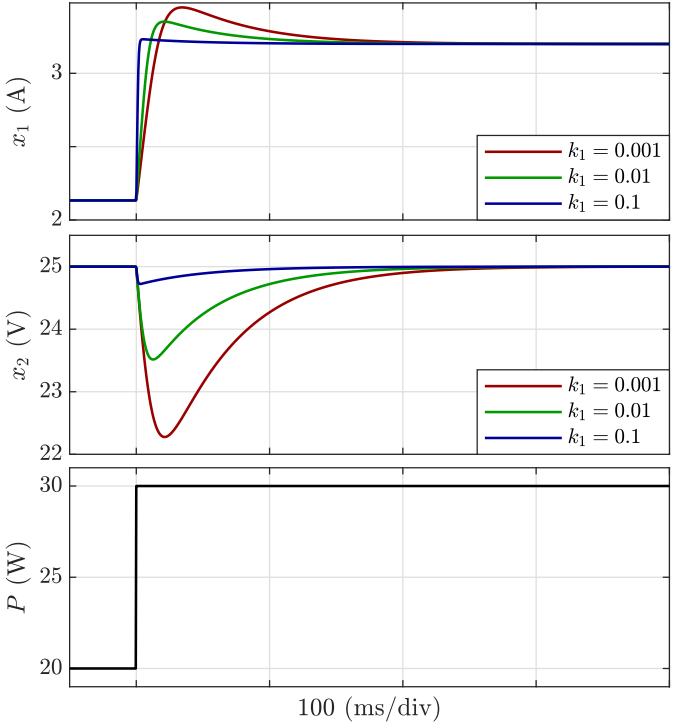}
\caption{Simulated current (top), voltage (middle) and power (bottom) waveforms for the adaptive IDA--PBC with $\gamma=20$ and different values of $k_1$.}\label{controlgain}
\end{figure}

Fig. \ref{controlgain} shows the profiles of the output voltage and inductor current for the adaptive IDA--PBC for different values of the control gain $k_1$ and adaptation gain $\gamma=20$, while applying a step change in the extracted power $P$ from 20 to 30 W. As shown in the figure, a larger control gain $k_1$ causes the output voltage to recover faster when $P$ is changed. However, for all values of $k_1$, the output voltage always converges to the desired equilibrium. This is due to the fact that, as predicted by the theory, the power estimated converges exponentially fast to the true value independently of the control signal.

\begin{figure}[!ht]
\centering
\includegraphics[width=\columnwidth]{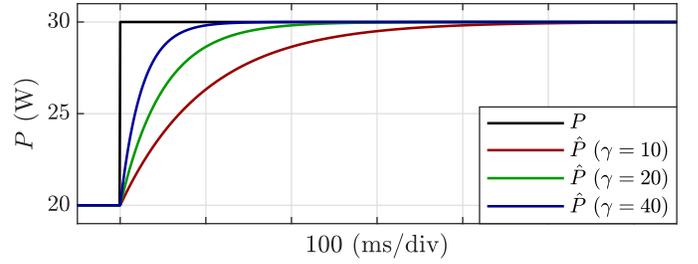}
\caption{Simulated transient performance of the estimate $\hat P$ under step changes of the parameter $P$ with $k_1=0.1$ and various estimator gains $\gamma$.}\label{estimator}
\vspace{-0.2cm}
\end{figure}

A step change in $P$ with its corresponding estimate $\hat P$ for different values of $\gamma$, is shown in Fig. \ref{estimator}. As predicted by the theory, for a larger $\gamma$, the speed of convergence of the estimator is faster. Notice, however, that in the selection of $\gamma$, there is a tradeoff between convergence speed and noise sensitivity.

\subsubsection{Phase plots}
Given that the IDA--PBC lives in the plane, it is possible to obtain a global picture of the behavior of these controllers by drawing their phase plot. These plots are obtained by performing averaged simulations of the closed loop system with a wide range of initial conditions for both states. For these simulations, it is assumed that $P$ is correctly estimated and $k_1$ is kept constant at a preselected value.

\begin{figure}[!ht]
\centering
\includegraphics[scale=0.80]{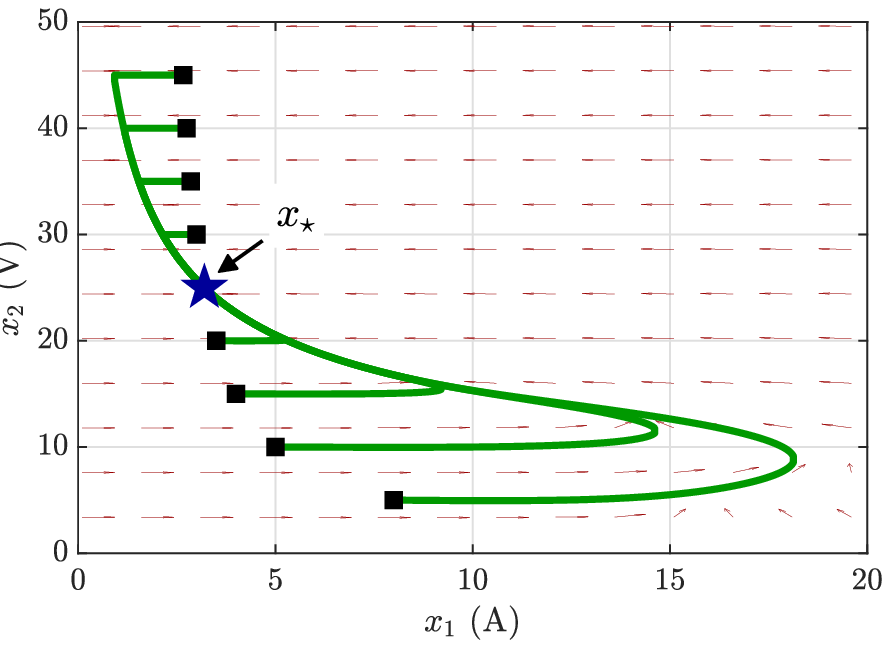}
\includegraphics[scale=0.80]{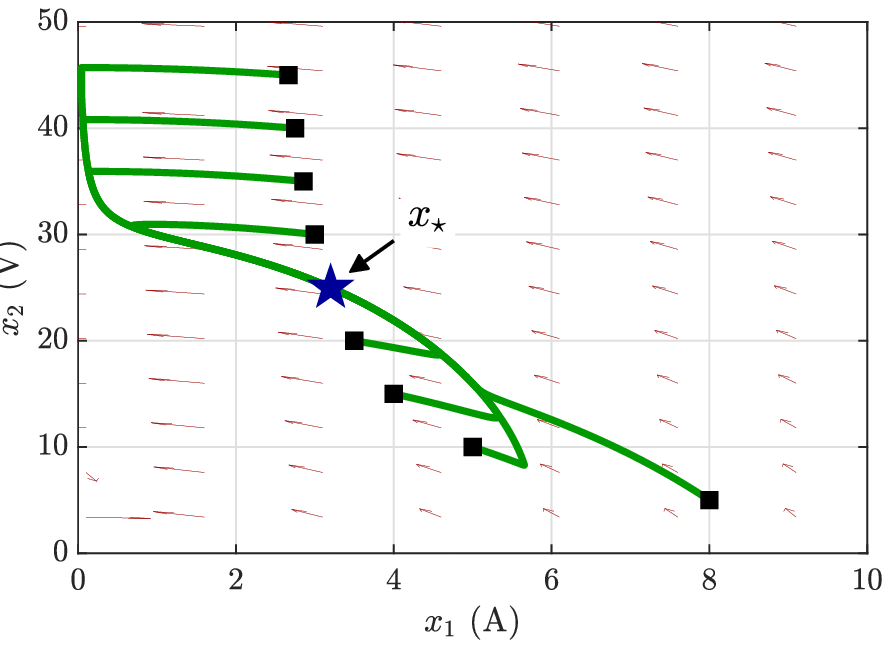}
\caption{Phase plots of the system with the IDA--PBC for different initial conditions and $k_1=0.001$.}
\label{fig:phase}
\vspace{-0.2cm}
\end{figure}


The transient behavior depicted in the phase plots depends greatly on the values of parameters $L$ and $C$. Given that the converter will be operating within a specified range for gain and throughput power, it can be designed so that the current and voltage ripples due to the switching action in $x_1$ and $x_2$, stay within pre--specified limits. For an example design with $E=15$ V, $P=30$ W and $x_{2\star}=25$ V, the equilibrium is defined as $x_\star=(x_{1\star},x_{2\star})=(3.2,25)$. Considering a switching frequency of 100 kHz and following the standard approach with ripples of 5\% and 1\% for $x_1$ and $x_2$ respectively, the minimum required values for $L$ and $C$ are found to be 5.859 mH and 480 $\mu$F, respectively. Fig. \ref{fig:phase} show the phase plots of the IDA--PBC together with some trajectories for different initial conditions and $k_1=0.001$. Fig. \ref{fig:phase} shows the phase plots for the system with the parameters of Table \ref{tab:simpars} in the top and in the bottom the phase plots for the example design. The initial conditions chosen ($\blacksquare$) represent equilibrium points with the same $P$ and different values for $x_{2\star}$. It can be seen that the system converges to the desired equilibrium point ({\color{myblue}$\bigstar$}) for a wide range of initial conditions $5~\mathrm{V} \leq x_2(0) \leq 45~\mathrm{V} $.  However, it is possible to show that the closed--loop vector field has another equilibrium in $\rea_{>0}^2$ that corresponds to a saddle point. Additionally, in can be observed that when $x_2(0)< x_{2\star}$, $x_1$ will have an overshoot that will increase as $x_2(0)$ approaches zero. However, proper selection of reactive elements can reduce this overshoot. Finally, it was observed that even if the values of $L$ and $C$ change but the ratio $L/C$ is kept constant, the trajectories followed by the system are the same.

{\color{black}\subsubsection{Comparison against a PI controller}
Herein, a comparison of the system in closed loop with the proposed IDA--PBC and a PI controller is presented. The converter is operated in boost mode with $E=15$ V and $x_{2_\star}=25$ V. A step change is applied to the power demanded by the CPL from 20 to 25 W. The PI controller seeks to drive to zero the error between the reference and measured values of the output voltage $x_2$.   The gains for the PI controller are chosen as $k_\mathrm{p}=0.002$ and $k_\mathrm{i}=0.001$. For the IDA--PBC, the gains are chosen as $\gamma=20$ and $k_1=0.3$. Fig. \ref{fig:cmp_1} illustrates the dynamic performance of the output voltage with both controllers.

\begin{figure}[!ht]
  \centering
  \includegraphics[scale=0.9]{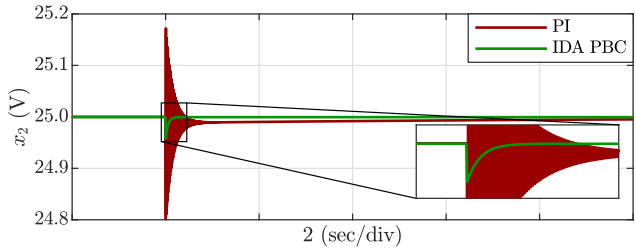}
  \caption{Output voltage $x_2$, contrasting the response of a PI controller against the proposed IDA--PBC.}\label{fig:cmp_1}
\end{figure}

It can be observed immediately that after the occurrence of the dynamic event the PI controller leads to an oscillatory response, and even after 8 s, it is still unable to achieve zero steady state error. Conversely, the proposed IDA--PBC takes less than 300 ms to recover from the step change in the load and does it with a much smoother dynamics. It should be noticed that the closed--loop system with traditional PI controller will be unstable under bigger variations of power load, which shows the poor robustness performance against larger disturbances. However, this problem does not exist in the proposed method since it is based on larger signal analysis.
}
%
%

\subsection{Switched simulation and experiments}

\begin{figure}[!ht]
  \centering\includegraphics[trim={0cm 0cm 0cm 0cm},clip,width=\columnwidth]{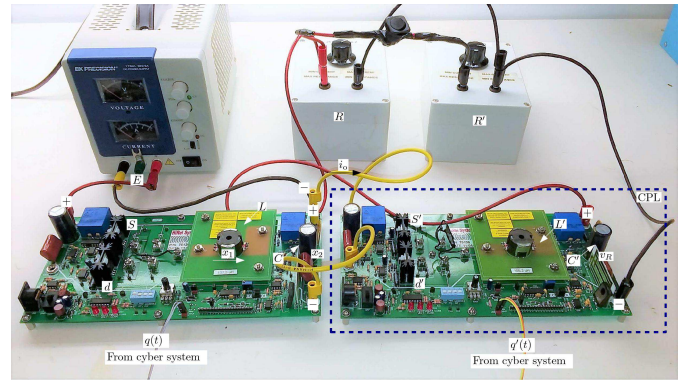}
  \caption{Experimental setup of the buck--boost converter feeding a CPL.}\label{fig:experiment}
\end{figure}

\ro{The realization of the buck--boost converter feeding a CPL (physical system) and its associated control scheme (control system) is illustrated in the schematic of Fig. \ref{block}. Dashed arrows are utilized to represent the flow of signals between the physical and control system. The emulation and control of the CLP is boxed using red dashed--dotted lines to highlight the fact that, while needed for the implementation, this part of system is not part of the proposed IDA--PBC.

\begin{figure}[!ht]
  \centering
  \includegraphics[scale=0.55]{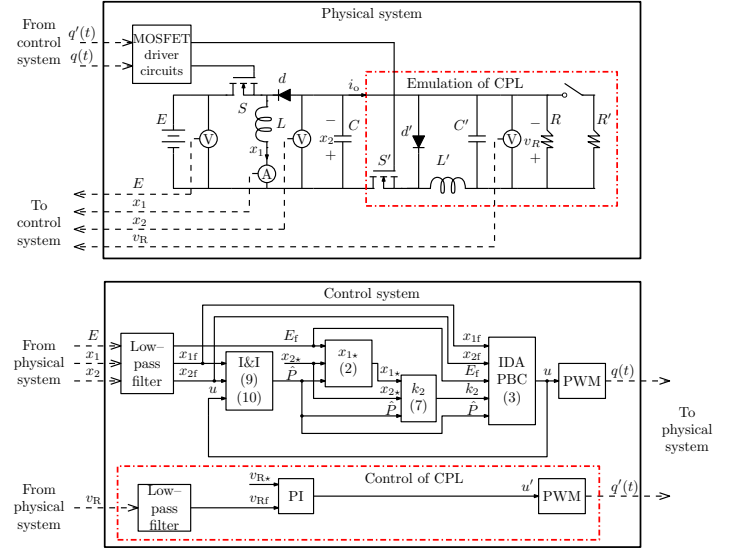}
  \caption{Schematic and control structure of a buck--boost converter feeding a CPL.}\label{block}
  \vspace{-0.2cm}
\end{figure}

The CPL is emulated via a tightly voltage--controlled buck converter (also utilizing a Vishay Dale board) and a resistor bank that can be switched in stages \cite{Rah09,kwasinski07,arora16}. It is also noted that within the control system an $\mathrm{f}$ appended to the subindex of a variable indicates that it has been filtered. This is done to convert switched signals into average ones. Furthermore, several blocks in the diagram point the equation number that is required to implement the corresponding action.} A picture of the experimental setup is shown in Fig. \ref{fig:experiment}.

For the experimental setup, the controller proposed in \eqref{ubucboo} with $k_1=0.01$, the I\&I estimator with $\gamma=20$ and the PI regulator for the CPL are implemented in a Texas Instrument DSP. The optimized floating--point math function library for this DSP is used, which allows for considerably faster execution speeds when performing tasks such as calculating trigonometric functions. Given that the controller is designed for the average model of the converter, the DSP samples the measured states at 10 kHz and then applies a low pass filter with a cutoff frequency of 1 kHz before they are fed to the controllers. For plotting purposes, signals from the experiment are acquired with an oscilloscope at a frequency of 50 kHz.

\begin{figure*}[!ht]
  \centering
  \includegraphics[width=\textwidth]{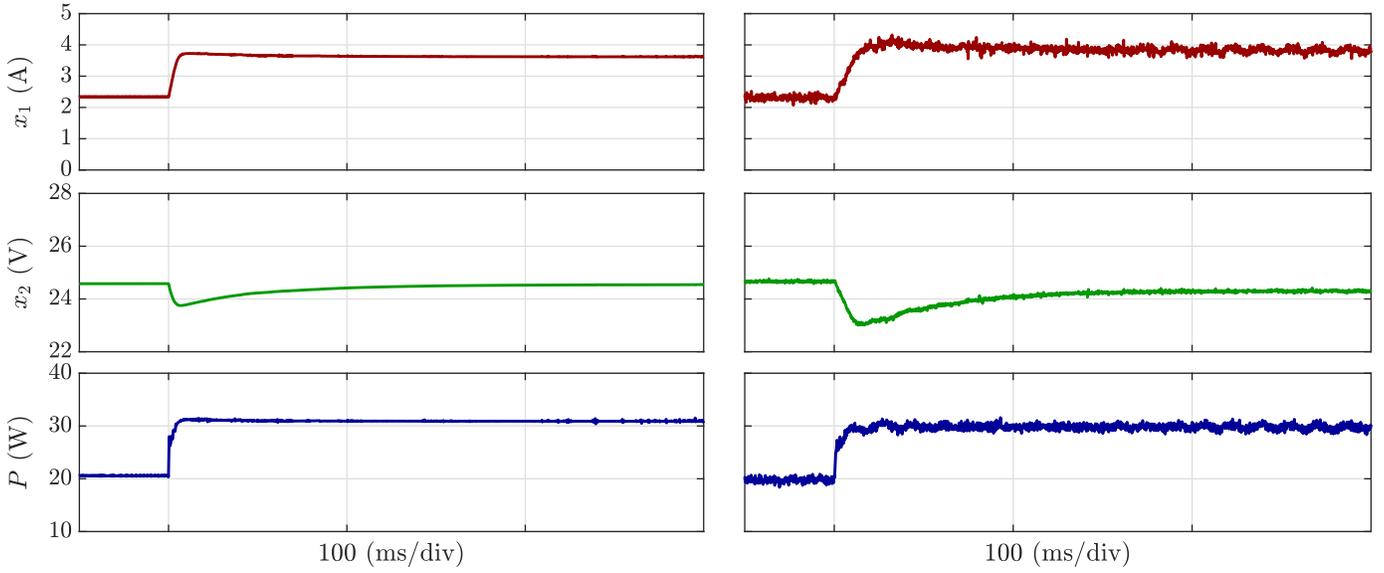}
  \caption{Simulated (left) and experimental (right) waveforms. Boost mode ($E=15$ V, $x_{2_\star}=25$ V) and $P$ stepped from $20$ W to $30$ W.}\label{exp1}
\end{figure*}

\begin{figure*}[!ht]
  \centering
  \includegraphics[width=\textwidth]{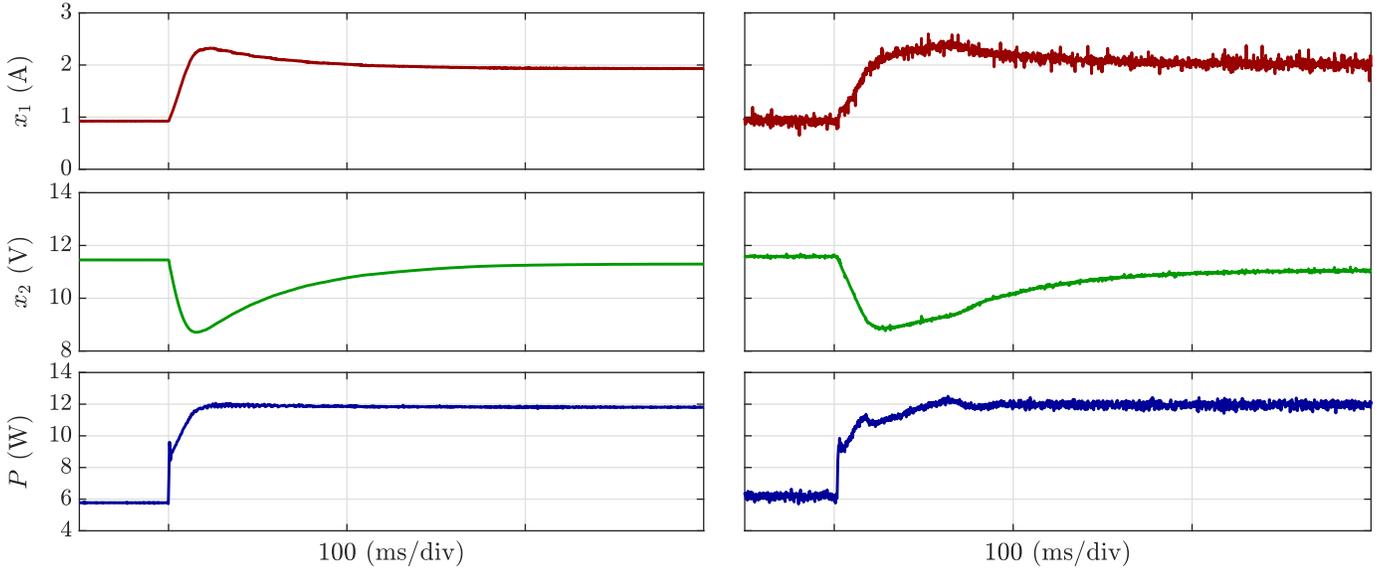}
  \caption{Simulated (left) and experimental (right) waveforms. Buck mode ($E=15$ V, $x_{2_\star}=12$ V) and $P$ stepped from $6$ W to $12$ W.}\label{exp3}
\end{figure*}

Given that the simulation uses a variable--step solver, the simulation results are resampled to 50 kHz. Both experimental and simulation waveforms are passed through a low pass filter with $f_c=1$ kHz before being plotted next to each other. The output current $i_\mathrm{o}$ is also measured and filtered, and then used to calculate the power shown in the results as $P=x_2i_\mathrm{o}$.

\subsubsection{Step changes in $P$}
The ability of the controller to regulate the output voltage while the load changes is tested experimentally. The results of these experiments are compared against results of switched simulations. The switches and passive elements in the simulation are realized using components from the Simscape/SimPowerSystems library. For both simulations and experiments, the switching frequency is chosen at 75 kHz.

To validate the effectiveness of the proposed approach, two cases which represent different scenarios of interest in practical applications are presented. The first experiment validates the proposed control when the converter is operating in boost mode. The input voltage $E$ and desired output voltage $x_{2_\star}$ are set to $15$ V, $25$ V, respectively, while the load power $P$ is changed from $20$ W to $30$ W. It can be seen in Fig. \ref{exp1} that the output voltage and the inductor current settle to their stationary values with a good transient performance.

One additional experiment is carried out to examine the output voltage regulation when operating in buck mode. The input voltage $E$ and desired output voltage $x_{2_\star}$ are set to $15$ V and $12$ V, respectively. The load power is initially set to $P=6$ W and is increased to $12$ W. The resulting output voltage and inductor current are shown in Fig. \ref{exp3}.

Although the output voltage contains steady state errors caused by parasitic elements not considered in the ideal model for both operating modes, the proposed controller successfully regulates the voltage at the desired value, regardless of the changes in $P$.

\subsubsection{Line and load regulation}
\begin{figure}[!ht]
  \centering\includegraphics[width=\columnwidth]{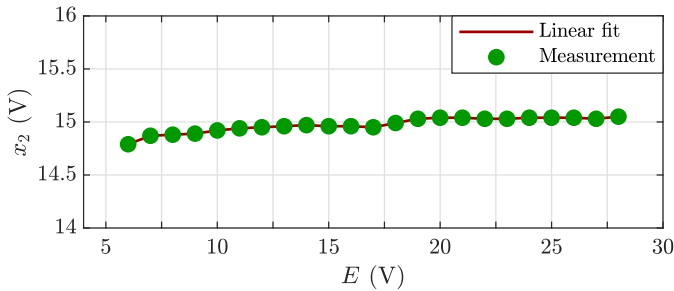}
  \caption{Line regulation plot. Output voltage against input voltage.}\label{fig:line_reg}
\end{figure}

The experimental ability of the proposed IDA--PBC to control the voltage under standard line/load regulation tests is presented herein. For line regulation, $x_{2_\star}=15$ V and $P=10$ W, while $E$ is being changed from $6$ to $28$ V in steps of $1$ V. It is noted that for this experiment the converter operating mode transitions from boost ($E < x_{2_\star}$) to buck ($E > x_{2_\star}$), and therefore the regulation plot is a composite of both operating modes. Fig. \ref{fig:line_reg} illustrates the results.

\begin{figure}[!ht]
  \centering
  \includegraphics[width=\columnwidth]{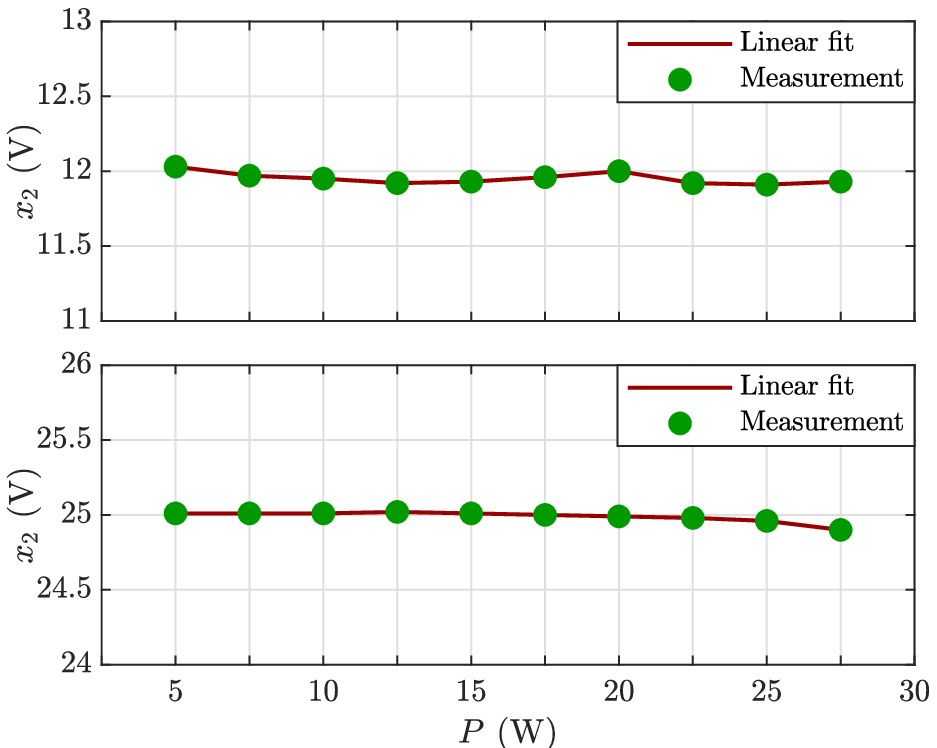}
  \caption{Load regulation plots for buck (top)  and boost (bottom) operating modes. Output voltage against load power.}\label{fig:load_reg}
\end{figure}

For load regulation, the input voltage is set to $E=15$ V and $x_{2_\star}=12$ V (buck mode), while the load was changed from $5~$W to $27.5~$W in steps of $2.5~$W. For boost mode, the input voltage is set to $E=15$ V and $x_{2_\star}=25$ V. The same values for $P$ were used as those in the buck mode experiment. Results are summarized in Fig. \ref{fig:load_reg}.%

\ro{It is readily observed from Fig. \ref{fig:line_reg} and Fig. \ref{fig:load_reg} that, despite small steady sate errors caused by the parasitic elements not considered in the mathematical model, the controller successfully regulates the output voltage in a hardware environment, and under relatively large variations in the input voltage and the output power. This indicates that the sophisticated IDA--PBC controller proposed herein will be suitable to be implemented in actual industrial applications.}

\section{Conclusions}\label{section5}

\ro{This paper has proposed a novel approach based on PBC to regulate the output voltage of DC--DC buck--boost converters feeding an unknown CPL. The control scheme assumes first that the CPL's power is known, and synthesizes an IDA--PBC that stabilizes the output voltage. Subsequently, an on--line I$\&$I estimator with global convergence has been presented to render the overall scheme adaptive, preserving asymptotic stability. The theoretical claims have been thoroughly validated via computer simulations and experimental prototyping, demonstrating the practical viability of the approach.}

\bibliographystyle{IEEEtran}
\bibliography{CPBC_bib}

\begin{thebibliography}{10}
\providecommand{\url}[1]{#1}
\csname url@samestyle\endcsname
\providecommand{\newblock}{\relax}
\providecommand{\bibinfo}[2]{#2}
\providecommand{\BIBentrySTDinterwordspacing}{\spaceskip=0pt\relax}
\providecommand{\BIBentryALTinterwordstretchfactor}{4}
\providecommand{\BIBentryALTinterwordspacing}{\spaceskip=\fontdimen2\font plus
\BIBentryALTinterwordstretchfactor\fontdimen3\font minus
  \fontdimen4\font\relax}
\providecommand{\BIBforeignlanguage}[2]{{%
\expandafter\ifx\csname l@#1\endcsname\relax
\typeout{** WARNING: IEEEtran.bst: No hyphenation pattern has been}%
\typeout{** loaded for the language `#1'. Using the pattern for}%
\typeout{** the default language instead.}%
\else
\language=\csname l@#1\endcsname
\fi
#2}}
\providecommand{\BIBdecl}{\relax}
\BIBdecl

\bibitem{khalig08}
A.~Khaligh, A.~M. Rahimi, and A.~Emadi, ``{Modified pulse--adjustment technique
  to control DC/DC converters driving variable constant--power loads},''
  \emph{IEEE Trans. Ind. Electron.}, vol.~55, no.~3, pp. 1133{--}1146, Mar.
  2008.

\bibitem{lefeuvre07}
E.~Lefeuvre, D.~Audigier, C.~Richard, and D.~Guyomar, ``{Buck--boost converter
  for sensorless power optimization of piezoelectric energy harvester},''
  \emph{IEEE Trans. Power Electron.}, vol.~22, no.~5, pp. 2018{--}2025, Sep.
  2007.

\bibitem{pavlovic14}
T.~Pavlovic, T.~Bjazic, and Z.~Ban, ``{Simplified averaged models of DC--DC
  power converters suitable for controller design and microgrid simulation},''
  \emph{IEEE Trans. Power Electron.}, vol.~28, no.~7, pp. 3266{--}3275, Jul.
  2013.

\bibitem{martin15}
A.~D. Martin, J.~M. Cano, J.~F.~A. Silva, and J.~R. Vázquez, ``{Backstepping
  control of smart grid{--}connected distributed photovoltaic power supplies
  for telecom equipment},'' \emph{IEEE Trans. Energy Convers.}, vol.~30, no.~4,
  pp. 1496{--}1504, Dec. 2015.

\bibitem{kwa11}
A.~Kwasinski and C.~N. Onwuchekwa, ``{Dynamic behavior and stabilization of DC
  microgrids with instantaneous constant--power loads},'' \emph{IEEE Trans.
  Power Electron.}, vol.~26, no.~3, pp. 822--834, Mar. 2011.

\bibitem{kar17}
M.~K. Zadeh, R.~Gavagsaz{--}Ghoachani, J.~P. Martin, S.~Pierfederici,
  B.~Nahid{--}Mobarakeh, and M.~Molinas, ``{Discrete--time tool for stability
  analysis of DC power electronics--based cascaded systems},'' \emph{IEEE
  Trans. Power Electron.}, vol.~32, no.~1, pp. 652--667, Jan. 2017.

\bibitem{Solsona15}
J.~A. Solsona, S.~G. Jorge, and C.~A. Busada, ``{Nonlinear control of a buck
  converter which feeds a constant power load},'' \emph{IEEE Trans. Power
  Electron.}, vol.~30, no.~12, pp. 7193--7201, Dec. 2015.

\bibitem{Du13}
W.~J. Du, J.~M. Zhang, Y.~Zhang, and Z.~M. Qiao, ``{Stability criterion for
  cascaded system with constant power load},'' \emph{IEEE Trans. Power
  Electron.}, vol.~28, no.~4, pp. 1843--1851, Apr. 2013.

\bibitem{geyer08}
T.~Geyer, G.~Papafotiou, and M.~Morari, ``Hybrid model predictive control of
  the step{--}down {DC--DC} converter,'' \emph{IEEE Trans. Control Syst.
  Technol.}, vol.~16, no.~6, pp. 1112{--}1124, Nov. 2008.

\bibitem{son12}
Y.~I. Son and I.~H. Kim, ``Complementary {PID} controller to {Passivity--Based}
  nonlinear control of boost converters with inductor resistance,'' \emph{IEEE
  Trans. Control Syst. Technol.}, vol.~20, no.~3, pp. 826{--}834, May 2012.

\bibitem{kim14}
S.~Kim, C.~R. Park, J.~Kim, and Y.~I. Lee, ``A stabilizing model predictive
  controller for voltage regulation of a {DC/DC} boost converter,'' \emph{IEEE
  Trans. Control Syst. Technol.}, vol.~22, no.~5, pp. 2016--2023, Sep. 2014.

\bibitem{wang16}
L.~Wang, Q.~H. Wu, Y.~K. Tao, and W.~H. Tang, ``Switching control of buck
  converter based on energy conservation principle,'' \emph{IEEE Trans. Control
  Syst. Technol.}, vol.~24, no.~5, pp. 1779{--}1787, Sep. 2016.

\bibitem{wang17}
J.~Wang, C.~Zhang, S.~Li, J.~Yang, and Q.~Li, ``Finite{--}time output feedback
  control for {PWM--}based {DC--DC} buck power converters of current sensorless
  mode,'' \emph{IEEE Trans. Control Syst. Technol.}, vol.~25, no.~4, pp.
  1359{--}1371, Jul. 2017.

\bibitem{Bar16}
N.~Barabanov, R.~Ortega, R.~Griñó, and B.~Polyak, ``{On existence and
  stability of equilibria of linear time--invariant systems with constant power
  loads},'' \emph{IEEE Trans. Circuits Syst. I, Reg. Papers}, vol.~63, no.~1,
  pp. 114--121, Jan. 2016.

\bibitem{Mar12}
D.~Marx, P.~Magne, B.~Nahid{--}Mobarakeh, S.~Pierfederici, and B.~Davat,
  ``{Large signal stability analysis tools in DC power systems with constant
  power loads and variable power loads---A review},'' \emph{IEEE Trans. Power
  Electron.}, vol.~27, no.~4, pp. 1773--1787, Apr. 2012.

\bibitem{Sin17}
S.~Singh, A.~R. Gautam, and D.~Fulwani, ``{Constant power loads and their
  effects in DC distributed power systems: A review},'' \emph{Renewable and
  Sustainable Energy Reviews}, vol.~72, pp. 407--421, May 2017.

\bibitem{emadi06}
A.~Emadi, A.~Khaligh, C.~H. Rivetta, and G.~A. Williamson, ``{Constant power
  loads and negative impedance instability in automotive systems: Definition,
  modeling, stability, and control of power electronic converters and motor
  drives},'' \emph{IEEE Trans. Veh. Technol.}, vol.~55, no.~4, pp. 1112--1125,
  Jul. 2006.

\bibitem{konst13}
G.~C. Konstantopoulos and A.~T. Alexandridis, ``Generalized nonlinear
  stabilizing controllers for {Hamiltonian--passive} systems with switching
  devices,'' \emph{IEEE Trans. Control Syst. Technol.}, vol.~21, no.~4, pp.
  1479{--}1488, Jul. 2013.

\bibitem{meshram18}
R.~V. Meshram, M.~Bhagwat, S.~Khade, S.~R. Wagh, A.~M. Stankovic, and N.~M.
  Singh, ``{Port--controlled phasor Hamiltonian modeling and IDA--PBC control
  of solid--state transformer},'' \emph{IEEE Trans. Control Syst. Technol.},
  pp. 1{--}14, 2018.

\bibitem{Cespedes11}
M.~Cespedes, L.~Xing, and J.~Sun, ``{Constant--power load system stabilization
  by passive damping},'' \emph{IEEE Trans. Power Electron.}, vol.~26, no.~7,
  pp. 1832--1836, Jul. 2011.

\bibitem{Rah10}
A.~M. Rahimi, G.~A. Williamson, and A.~Emadi, ``{Loop--cancellation technique:
  A novel nonlinear feedback to overcome the destabilizing effect of
  constant--power loads},'' \emph{IEEE Trans. Veh. Technol.}, vol.~59, no.~2,
  pp. 650--661, Feb. 2010.

\bibitem{Salimi16}
M.~Salimi and S.~Siami, ``Cascade nonlinear control of {DC}{--}{DC}
  {buck/boost} converter using exact feedback linearization,'' in \emph{2015
  4th Int. Conf. Elect. Power \& Energy Convers. Syst. (EPECS)}, Nov. 2015.

\bibitem{Rah09}
A.~M. Rahimi and A.~Emadi, ``{Active damping in DC/DC power electronic
  converters: A novel method to overcome the problems of constant power
  loads},'' \emph{IEEE Trans. Ind. Electron.}, vol.~56, no.~5, pp. 1428--1439,
  May 2009.

\bibitem{Sin16}
S.~Singh, N.~Rathore, and D.~Fulwani, ``{Mitigation of negative impedance
  instabilities in a {DC/DC} buck{--}boost converter with composite load},''
  \emph{Journal of Power Electron.}, vol.~16, no.~3, pp. 1046--1055, May 2016.

\bibitem{He2018}
W.~He, R.~Ortega, J.~E. Machado, and S.~H. Li, ``{An adaptive passivity--based
  controller of a buck--boost converter with a constant power load},''
  \emph{Asian Journal of Control}, vol.~21, no.~2, pp. 1{--}15, Mar. 2018.

\bibitem{Ortega02}
R.~Ortega, A.~V.~D. Schaft, B.~Maschke, , and G.~Escobard, ``{Interconnection
  and damping assignment passivity--based control of port--controlled
  Hamiltonian systems},'' \emph{Automatica}, vol.~38, no.~4, pp. 585--596, Apr.
  2002.

\bibitem{Ast08}
A.~Astolfi, D.~Karagiannis, and R.~Ortega, \emph{{Nonlinear and adaptive
  control with applications}}.\hskip 1em plus 0.5em minus 0.4em\relax Berlin:
  Springer{--}Verlag, 2008.

\bibitem{He2018CEP}
W.~He, C.~A. Soriano-Rangel, R.~Ortega, A.~Astolfi, F.~Mancilla-David, and
  S.~Li, ``{Energy shaping control for buck--boost converters with unknown
  constant power load},'' \emph{Control Engineering Practice}, vol.~74, pp.
  33{--}43, May 2018.

\bibitem{ACC2018}
W.~{He}, C.~{Soriano--Rangel}, R.~{Ortega}, A.~{Astolfi}, F.~{Mancilla--David},
  and S.~{Li}, ``{DC--DC} buck{--}boost converters with unknown {CPL}: An
  adaptive {PBC},'' in \emph{2018 Annual American Control Conference (ACC)},
  Jun. 2018, pp. 6749{--}6754.

\bibitem{Sep97}
R.~Sepulchre, M.~Jankovic, and K.~P., \emph{Constructive nonlinear
  control}.\hskip 1em plus 0.5em minus 0.4em\relax London: Springer{--}Verlag,
  1997.

\bibitem{kwasinski07}
A.~Kwasinski and P.~T. Krein, ``Stabilization of constant power loads in
  {DC}{--}{DC} converters using passivity{--}based control,'' in \emph{2007
  29th Int. Telecommun. Energy Conf. (INTELEC)}, Sep. 2007, pp. 867{--}874.

\bibitem{arora16}
S.~Arora, P.~T. Balsara, and D.~K. Bhatia, ``{Digital implementation of
  constant power load (CPL), active resistive load, constant current load and
  combinations},'' in \emph{2016 IEEE Dallas Circuits and Systems Conference
  (DCAS)}, Oct. 2016, pp. 1{--}4.

\end{thebibliography}

\appendix

\subsection{Proof of proposition \ref{proposition0}}
\label{app1}
It will be shown that the control \eqref{ubucboo} can be derived using the IDA--PBC method of \cite{Ortega02}  with the selection
\begin{align}
F_d(x):=\left[
\begin{array}{cc}
 -\frac{x_2}{Lx_1} & -\frac{2x_2}{Cv_\mathrm{T}} \\
 \frac{2x_2}{Cv_\mathrm{T}} & -\frac{2LEx_1}{C^2v_\mathrm{T}^2}\\
 \end{array}
 \right],
\end{align}
that, for $x \in \rea_{>0}^2$, satisfies the condition $F_d(x)+F_d(x)^T<0$.

The system \eqref{bucboo} can be rewritten in the form
$$\dot x=f(x)+g(x)u$$
where
\begequ
\lab{g}
g(x):=\left[
  \begin{array}{c}
    C(x_2+E) \\
    -Lx_1 \\
  \end{array}
\right]
\endequ
is the system input matrix, and
$$
f(x):=\left[
  \begin{array}{c}
    -\frac{x_2}{L} \\
    \frac{x_1}{C}-\frac{P}{Cx_2} \\
  \end{array}
\right]
$$
\begin{figure*}[!t]
\normalsize
\begin{align}\label{eq11}
   \left[
     \begin{array}{cc}
       Lx_1 & C(x_2+1) \\
     \end{array}
   \right]
\left(\left[
    \begin{array}{c}
      -\frac{x_2}{L} \\
      \frac{x_1}{C}-\frac{P}{Cx_2} \\
    \end{array}
  \right]
-\left[
   \begin{array}{cc}
 -\frac{x_2}{Lx_1} & -\frac{2x_2}{Cv_\mathrm{T}} \\
 \frac{2x_2}{Cv_\mathrm{T}} & -\frac{2LEx_1}{C^2v_\mathrm{T}^2}\\
 \end{array}
 \right]
\nabla H_d(x)
\right)=0,
\end{align}
\end{figure*}

is the vector field. Noting that the left annihilator of $g(x)$ is $g^\perp(x):=[Lx_1\;C(x_2+E)]$, the PDE takes the form of \eqref{eq11}, which is equivalent to
\begin{align}\label{pde1}
\scalebox{0.99}{$%
   -x_2\nabla_{x_1} H_d(x)+\frac{2Lx_1}{C}\nabla_{x_2} H_d(x)=P - Ex_1+\frac{PE}{x_2}.$}
\end{align}
\begin{figure*}[!t]
\normalsize
\begin{align}\label{eq12}
H_d(x)=-\frac{\sqrt{C}}{2L}\left(\sqrt{C}Ex_2+\sqrt{2L}P\arctan{\left[\frac{\sqrt{2L}x_1}{\sqrt{C}x_2}\right]}\right)
-\frac{\sqrt{2L}PE\arctan{\left[\frac{\sqrt{2L}x_1}{\sqrt{W}}\right]}}{\sqrt{\frac{W}{C}}} +\Phi\left(\frac{Lx_1^2}{C}+\frac{x_2^2}{2}\right).
\end{align}
\end{figure*}
The solution of \eqref{pde1} is easily obtained using a symbolic language, {\em e.g.}, Maple or Mathematica, and is of the form of \eqref{eq12}, where $\Phi(\cdot)$ is an arbitrary function. Selecting this free function as
$$
\Phi(z):=\frac{k_1}{2}(z+ k_2)^2,
$$
with $k_1$ and $k_2$ arbitrary constants, yields \eqref{solution}.

To complete the design it only remains to prove the existence of $k_1$ and $k_2$. Towards this end, the gradient is first computed as \eqref{eq13}
\begin{figure*}[!t]
\normalsize
\begin{align}\label{eq13}
\nabla H_d  &=\begin{bmatrix}
          \frac{W \left(C^3 P (E-x_2)+C^2 k_1 L x_1 x_2^2 \left(2 k_2+x_2^2\right)+4 C k_1 L^2 x_1^3 \left(k_2+x_2^2\right)+4 k_1 L^3 x_1^5\right)-\sqrt{2LC^6W} E P x_1 \arctan\left(\frac{\sqrt{2L}  x_1}{\sqrt{W}}\right)}{C^2 W^2}\\
          \frac{\sqrt{\frac{W}{C}} \left(-E C^3 x_2^3+C^2 L \left(2 k_1 k_2 x_2^4+k_1 x_2^6+2 P x_1 x_2-2 E x_1 (P+x_1 x_2)\right)+4 C k_1 L^2 x_1^2 x_2^2 \left(k_2+x_2^2\right)+4 k_1 L^3 x_1^4 x_2^2\right)-\sqrt{2LC^5}E P x_2^2 \arctan\left(\frac{\sqrt{2L}  x_1}{\sqrt{W}}\right)}{2L\sqrt{C}W^3}
          \end{bmatrix}
\end{align}
\hrulefill
\end{figure*}

Evaluating it at the equilibrium and selecting $k_2$ as given in \eqref{k2} yields
\begin{align}
\left.\nabla H_d\right|{x=x\star}=\begin{bmatrix}0\\
\frac{\sqrt{C} \left(\sqrt{C x_{2\star}^2} (P-E x_{1\star})+E \sqrt{C} P\right)}{2 L x_{1\star} x_{2\star}}\end{bmatrix}.
\end{align}
Invoking $x_{1\star}=P\left(\frac{1}{x_{2\star}}+\frac{1}{E} \right)$ one gets $\left.\nabla H_d\right|_{x=x\star}=0$.

On the other hand, the Hessian of $H_d(x)$ is given by
\begin{align}\label{hessian}
\nabla^2 H_d=\begin{bmatrix} \nabla_{x_1}^2 H_d
&\nabla_{x_1x_2}^2 H_d \\
\nabla_{x_2x_1}^2 H_d
&\nabla_{x_2}^2 H_d\end{bmatrix},
\end{align}
where the elements are given as in \eqref{eq:grad1}--\eqref{eq:grad3}.
\begin{figure*}[!t]
\normalsize
\begin{align}
\nabla_{x_1}^2 H_d=&\frac{1}{\sqrt{C^5 x_2^2} W^3}\Bigg(\sqrt{2L}C^4x_2 EP \sqrt{\frac{W}{C}} \left(C x_2^2-4 L x_1^2\right) \arctan\left(\frac{\sqrt{2L} x_1}{\sqrt{W}}\right) L W \bigg(4 \sqrt{C^7} P x_1 x_2^2 + \sqrt{C^7 x_2^2} \left(2 k_1 k_2 x_2^4 \right. \nonumber\\
& \left. +k_1 x_2^6+6 E P x_1\right)+4 L^2 \sqrt{C^3} k_1  x_1^4 x_2^2 \left(2 k_2+7 x_2^2\right)+(2 L \sqrt{C^5}k_1 x_1^2 x_2^3) \left(4 k_2+5x_2^2\right)+24 L^3 \sqrt{C} k_1 x_1^6 x_2^2\bigg)\Bigg) \label{eq:grad1} \\
\nabla_{x_2x_1}^2 H_d=&\frac{1}{C W^2 x_2}\Bigg(-2 L C^2 P x_1 x_2 -\frac{3 \sqrt{2LC^5} E P x_1 x_2^2 \arctan\left(\frac{\sqrt{2L} x_1}{\sqrt{W}}\right)}{\sqrt{\frac{W}{C}}}+C^{3} P x_2+ 2 C^3 E P x_2^2 +2 L C^2 x_1\nonumber\\
&  \left(k_1 x_2^6-E P x_1\right)+8 L^2 C k_1  x_1^3 x_2^4+8 L^3 k_1  x_1^5 x_2^2\Bigg) \label{eq:grad2} \\
\nabla_{x_2}^2 H_d=&\frac{1}{2 \sqrt{LC^3} x_2^3 W^3}\Bigg(2 \sqrt{2} C^4 E P x_2^3 \sqrt{\frac{W}{C}} \left(L x_1^2-C x_2^2\right) \arctan\left(\frac{\sqrt{2L} x_1}{\sqrt{W}}\right)+ \sqrt{L} W \bigg(-4 \sqrt{C^{7}} P x_1 x_2^4 \nonumber\\
& +2 L \sqrt{C^5} x_1^2 x_2^3 \left(4 k_1 k_2 x_2^4+7 k_1 x_2^6-2 E P x_1\right)+ \sqrt{C^7}x_2 \big(2 k_1 k_2 x_2^4+3 k_1x_2^6 -8 E P x_1\big) \nonumber\\
&+4 k_1 L^2 \sqrt{C^3} x_1^4 x_2^3 \left(2 k_2+5 x_2^2\right)+8 k_1 L^3 x_1^6 x_2^2 \sqrt{C x_2^2}\bigg)\Bigg) \label{eq:grad3}.
\end{align}
\hrulefill
\end{figure*}
Replacing $k_2$ in \eqref{hessian} and evaluating it at the equilibrium point $x=x_\star$, it follows
\begin{align}\label{hessian1}
\nabla^2 H_d\Bigg|_{x=x\star}=\begin{bmatrix} \nabla_{x_1}^2 H_d|_{x=x\star}
&\nabla_{x_1x_2}^2 H_d|_{x=x\star}\\
\nabla_{x_2x_1}^2 H_d|_{x=x\star}
&\nabla_{x_2}^2 H_d|_{x=x\star}\end{bmatrix},
\end{align}
where the elements are given as in \eqref{eq:gradeq1}--\eqref{eq:gradeq3}.
\begin{figure*}[!t]
\normalsize
\begin{align}
\nabla_{x_1}^2 H_d|_{x=x\star}=&\frac{1}{C^2 W_\star^3 x_{1\star} }\Bigg(W_\star \bigg(C^4 P x_{2\star}^2 (x_{2\star}+E)+2 L C^3 P x_{1\star}^2 (3 x_{2\star}+4 E)+4 L^2 C^2 k_1  x_{1\star}^3 x_{2\star}^4+\nonumber\\
&16 L^3 C k_1  x_{1\star}^5 x_{2\star}^2+16 L^4 k_1  x_{1\star}^7\bigg)-6 \sqrt{2L^3C^7} E P x_{1\star}^3 \sqrt{\frac{W_\star}{C}} \arctan\left(\frac{\sqrt{2L} x_{1\star}}{\sqrt{W_\star}}\right)\Bigg)\label{eq:gradeq1} \\
\nabla_{x_2x_1}^2 H_d|_{x=x\star}=&\frac{1}{\sqrt{CW^5}x_{2\star}}\Bigg(\sqrt{\frac{W}{C}} \Big(C^3 Px_{2\star}^2 (x_{2\star}+2E)-2 L C^2 x_{1\star} \bigg(-k_{1} x_{2\star}^6+P x_{1\star} x_{2\star} +E P x_{1\star}\bigg)\nonumber\\
&+8 L^2 C k_{1}  x_{1\star}^3 x_{2\star}^4+8 L^3 k_{1}  x_{1\star}^5 x_{2\star}^2\Big)-3 \sqrt{2LC^5} E P x_{1\star} x_{2\star}^2 \arctan\left(\frac{\sqrt{2L} x_{1\star}}{\sqrt{W_\star}}\right)\Bigg)\label{eq:gradeq2} \\
\nabla_{x_2}^2 H_d|_{x=x\star}&=\frac{1}{2\sqrt{3CW^5}x_{2\star}^2}\Bigg(2 \sqrt{2C^5} E P x_{2\star}^2 \left(L x_{1\star}^2-C x_{2\star}^2\right) \arctan\left(\frac{\sqrt{2L} x_{1\star}}{\sqrt{W_\star}}\right)+ \sqrt{\frac{L W_\star}{C}}\nonumber\\
&  \Big(C^3 x_{2\star}^2 \left(2 k_1 k_2 x_{2\star}^4+3k_1x_{2\star}^6-4 P x_{1\star} x_{2\star}-8 E P x_{1\star}\right)+2 L C^2 x_{1\star}^2 \big(4 k_1 k_2 x_{2\star}^4+7 k_1 x_{2\star}^6\nonumber\\
&-2 E P x_{1\star}\big)+4 L^2 C k_1  x_{1\star}^4 x_{2\star}^2 \left(2 k_2+5x_{2\star}^2\right)+8 k_1 L^3 x_{1\star}^6 x_{2\star}^2\Big)\Bigg) \label{eq:gradeq3}.
\end{align}
\hrulefill
\begin{align}\label{eq16}
k_1'&:=-\frac{C^3 P \left(2 \sqrt{L} x_{1\star} (2 x_{2\star}+3 E) W_\star+\sqrt{2C} E  \sqrt{\frac{W_\star}{C}} \left(C x_{2\star}^2-4 L x_{1\star}^2\right) \arctan\left(\frac{\sqrt{2L} x_{1\star}}{\sqrt{W_\star}}\right)\right)}{\sqrt{L}W_\star^3 \left(C \left(2 k_2+x_{2\star}^2\right)+6 L x_{1\star}^2\right)},
\end{align}
\begin{align}\label{eq19}
k_1''&:=\frac{1}{2 L \sqrt{\frac{W^5}{C}} x_{1\star}  \bigg(C x_{2\star}^3+E C x_{2\star}^2-2 E L x_{1\star}^2\bigg)}\Bigg(3 \sqrt{2LC^5} E P x_{1\star} \bigg(C x_{2\star}^3+E C x_{2\star}^2 -2 E L x_{1\star}^2\bigg) \nonumber\\
& \arctan\left(\frac{\sqrt{2L} x_{1\star}}{\sqrt{W_\star}}\right)-\sqrt{C^5 W} P  \bigg(2 E x_{2\star} \left(C x_{2\star}^2-5 L x_{1\star}^2\right)+E^2 \left(C x_{2\star}^2-10 L x_{1\star}^2\right)+\nonumber\\
&C x_{2\star}^4-2 L x_{1\star}^2 x_{2\star}^2\bigg)\Bigg).
\end{align}
\end{figure*}

Some lengthy, but straightforward, calculations prove that $\nabla_{x_1}^2 H_d|{x=x\star}>0$ holds if and only if $k_1 > k_1'$ where $k_1'$ is defined as in \eqref{eq16} and $k_1''$ is defined as in \eqref{eq19}.

Finally, in order to ensure $\det\left(\left.\nabla^2 H\right|{x=x\star}\right)>0$, $k_1$ should be chosen such that $k_1> \max {\{k_1', k_1''\}}$. This ensures $\nabla^2 H|{x=x\star}>0$, which ends the proof that $x_\star$ is an asymptotically stable equilibrium of the closed--loop.

The proof of the existence of an estimate of the domain of attraction follows immediately noting that it has been shown above that the function $H_d(x)$ has a positive definite Hessian evaluated at $x_\star$, therefore it is {\em convex}. Consequently, for sufficiently small $c$, the sublevel set $\Omega_x$ defined in \eqref{region} is bounded and strictly contained in $\mathbb{R}^2_{>0}$. The proof is completed recalling  that sublevel sets of strict Lyapunov functions are inside the domain of attraction of the equilibrium.

\qed

\subsection{Proof of proposition \ref{proposition2}}
\label{app2}
Differentiating $\tilde P$ along the trajectories of \eqref{bucboo} and using \eqref{boest1} one gets
\begin{align}
\dot {\tilde P}
&= -\gamma x_2C\dot x_2 + \dot {P}_I\nonumber\\
&= -\gamma x_1Cx_2(1-u) + \gamma P + \dot {P}_I.\nonumber
\end{align}
Substituting \eqref{boest2} in the last equation yields
\begin{align}
\dot {\tilde P}&= \gamma P + \frac{1}{2}\gamma^2Cx_2^2 -\gamma P_I\nonumber\\
&=-\gamma \tilde P,\nonumber
\end{align}
which reveals that the estimation $\hat P$ will exponentially converge to $P$.

\qed

\end{document}